\documentclass[a4paper,11pt]{article}
\usepackage{graphicx,rotating,slashed,amsmath,xcolor,amssymb,amsfonts,colortbl,cite,subcaption}
\usepackage[totalheight = 23cm, totalwidth = 17cm]{geometry}
\usepackage{physics}
\usepackage{jheppub,amscd,braket,amsthm}
\usepackage{bm}
\numberwithin{equation}{section}

\definecolor{bluc}{cmyk}{1,1,0,0.1}
\definecolor{rossoCP3}{cmyk}{0,.88,.77,.40}
\definecolor{rosso}{cmyk}{0,1,1,0.4}
\definecolor{rossos}{cmyk}{0,1,1,0.55}
\definecolor{rossoc}{cmyk}{0,1,1,0.2}
\definecolor{verdes}{cmyk}{0.92,0,0.59,0.4}

\hypersetup{colorlinks, bookmarksopen, bookmarksnumbered,
citecolor=verdes, linkcolor=bluc, pdfstartview=FitH, urlcolor=rossos}

\usepackage{comment}

\usepackage{mdframed}

\newcommand{\mpl}{m_{\rm Pl}}

\newcommand{\bea}{\begin{eqnarray}}
\newcommand{\eea}{\end{eqnarray}}
\newcommand{\be}{\begin{equation}}
\newcommand{\ee}{\end{equation}}

\definecolor{purpleheart}{rgb}{0.41, 0.21, 0.61}

\definecolor{limegreen}{rgb}{0.2, 0.8, 0.2}

\newcommand{\ud}{\,\mathrm{d}} 

\preprint{APCTP Pre2025-025}

\title{Gravitational open effective field theory of inflation}

\author[a]{Perseas Christodoulidis,}
\author[a,b]{and Jinn-Ouk Gong}
\affiliation[a]{Department of Science Education, Ewha Womans University,\\ 52 Ewhayeodae-gil, Seodaemun-gu, Seoul 03760, Korea}
\affiliation[b]{Asia Pacific Center for Theoretical Physics,\\ 77 Cheongam-ro, Nam-gu, Pohang 37673, Korea}
\emailAdd{perseas@ewha.ac.kr}
\emailAdd{jgong@ewha.ac.kr}

\abstract{
We construct a gravitational open extension of the effective field theory of inflation in the Schwinger–Keldysh framework. While physical symmetries allow many open operators in the Schwinger–Keldysh action, most of them overconstrain the equations of motion, yielding inconsistent dynamics. We identify a minimal open operator compatible with propagating scalar and tensor modes and build a local gravitational action, recovering known dissipative models of inflation.
}

\begin{document}

\maketitle
\flushbottom

\section{Introduction}

The Schwinger–Keldysh (SK) formalism provides a systematic way to describe quantum fields interacting with unobserved degrees of freedom which can lead to dissipation, the appearance of stochastic noise, and more generally to modified conservation laws \cite{Schwinger:1960qe,Keldysh:1964ud,Feynman:1963fq,Sieberer:2015hba,Crossley:2015evo,Glorioso:2017fpd,Liu:2018kfw,Haehl:2016pec}. This formalism has enabled the development of open effective field theories (EFTs), in which physical symmetries rather than a single-copy action determine the allowed dynamics. Recent work has applied these ideas in a wide range of settings, from systems with approximate global symmetries to open formulations of electromagnetism and gravity \cite{Hongo:2018ant,Delacretaz:2021qqu,Armas:2021vku,Baggioli:2023tlc,Salcedo:2024nex,Salcedo:2024smn,Lau:2024mqm,Salcedo:2025ezu,Christodoulidis:2025ymc,Kaplanek:2025moq}. Applying this framework to cosmology is particularly relevant: during inflation, the inflaton may couple to additional (heavy) fields or dark sectors whose effects cannot be captured by a closed EFT. Such interactions can modify the evolution of perturbations and lead to dissipative dynamics such as warm inflation~\cite{Berera:1995ie}. A consistent SK EFT formulation is therefore essential for understanding how these open-system effects can be incorporated into the inflationary dynamics in a systematic way while preserving their gravitational couplings.

In the SK framework for gravity, or more generally gauge theories, one often works directly at the level of the equation of motion (EOM), without an a priori guarantee that these equations are dynamically consistent. In closed theories, consistency is ensured by constructing the EFT at the level of the action. Noether’s second theorem then ensures the presence of identities among the EOM, effectively reducing their number to match the number of gauge-fixed variables.  This raises a fundamental question: how can we determine whether a given set of EOMs in open theories forms a consistent system? Although deformed (open-system) analogues of Noether identities have been identified in simple open extensions of gravity and electromagnetism, such identities are not expected to exist for the most general open terms that respect the physical symmetries of the system  \cite{Christodoulidis:2025ymc}.

In this work, we examine the open EFT of inflation in the Markovian approximation and identify a concrete open term that gives rise to deformed identities and, therefore, produces consistent dynamics. After solving the constraints, we show that the resulting theory propagates one scalar and two tensor degrees of freedom, providing a consistent local open extension of the closed EFT in the regime where environmental modes relax rapidly. We further derive the EOM for the curvature perturbation, recovering the form previously obtained in works that built the EFT for scalars only \cite{LopezNacir:2011kk,Creminelli:2023aly,Salcedo:2024smn}.

This work is organized as follows. In Section~\ref{sec:closed_eft} we present a recap of the standard EFT of inflation but written in the SK formalism. In Section~\ref{sec:consistent_eft} we demonstrate that only a subset of all possible open terms can produce an EFT with propagating degrees of freedom. In Section~\ref{sec:open_eft} we construct a local open EFT of inflation. Finally, in Section~\ref{sec:summary} we summarize our main findings.

\section{The closed EFT of inflation} 
\label{sec:closed_eft}

\subsection{Double metrics and diffeomorphisms}
To write an open SK EFT we start by doubling all fields of the theory. For gravity, we double the metric field $\{g_{1\mu \nu},g_{2\mu \nu} \}$, keeping in mind that the two copies of the spacetime are defined with respect to a reference spacetime via the pullback action and then rotate to the Keldysh basis:
\begin{align}
g_{\mu \nu} & \equiv {1 \over 2} (g_{1\mu \nu} + g_{2\mu \nu}) \, , 
\\
g_{\rm a \mu \nu} & \equiv g_{1\mu \nu} - g_{2\mu \nu} \, , 
\end{align}
where the subscript ``a'' stands for ``advanced'' metric. However, the naive expectation of the SK action for the Einstein-Hilbert action 
\be
S_{\rm naive} \overset{?}{=} \int \ud^4 x \sqrt{-g} G_{\mu \nu} F^{\mu \nu} (g_{\rm a}^{\kappa \lambda}) + \cdots
\ee
for some function of the advanced metric, does not seem very feasible due to the non-linearities of gravity. To alleviate this problem, we can restrict our attention to almost aligned configurations where the advanced metric can be treated as a fluctuation field living in the physical spacetime resulting in an action:
\be
S_{\rm SK} =  \int \ud^4 x \sqrt{-g} {\delta S \over \delta g^{\mu \nu}} g_{\rm a}^{\mu \nu} + \cdots \, .
\ee
The previous structure is reminiscent of the variation of the usual Einstein-Hilbert action
\be
\delta S =  \int \ud^4 x \sqrt{-g} {\delta S \over \delta g^{\mu \nu}} \delta g^{\mu \nu} \, ,
\ee
and so we can read the advanced symmetry of the non-dissipative theory 
\be
\delta g_{\rm a}^{\mu \nu} = \mathcal{L}_{\xi} g^{\mu \nu} \, ,
\ee
which is referred to as \textit{noise diffeomorphism} \cite{Liu:2018kfw}. This symmetry is equivalent to an off-shell relation between the EOMs (without using their solutions)
\be
\nabla^{\mu} {\delta S \over \delta g^{\mu \nu}} = 0 \, ,
\ee
which is a property of diffeomorphism invariant theories.

However, there is no transformation acting only on the physical metric that leaves the action invariant, in contrast to e.g.~the open EFT of Maxwell theory, where the retarded/advanced gauge fields are transformed independently. Here, physical diffeomorphisms act on every tensor of the theory, and so under those transformations, fields transform as
\be
\delta g_{\mu \nu} = - \mathcal{L}_{\xi} g_{\mu \nu}  \, , \quad \delta g_{\rm a \mu \nu} = - \mathcal{L}_{\xi} g_{\rm a \mu \nu} \, , \quad \text{etc}
\ee
assuming that the SK action is diffeomorphism invariant.

\subsection{Recap of the standard EFT of inflation}

To extend the simple Einstein-Hilbert term to the closed SK EFT of inflation, we simply use the EOMs of the standard EFT of inflation \cite{Cheung:2007st} plus the advanced metric. In the following, we review the minimal ingredients of the EFT of inflation when working in the unitary gauge where field fluctuations are set to zero ($\delta \phi =0)$. Imposing a time-dependent Friedmann–Lema\^{i}tre–Robertson–Walker (FLRW) background breaks time diffeomorphisms, and the action for fluctuations becomes invariant only under time-dependent spatial diffeomorphisms. The building blocks of the EFT include the following:
\begin{enumerate}

\item Functions of $t$.

\item The one-form $\ud t \equiv \delta^0_{\mu} \ud x^{\mu}$, which remains invariant under spatial reparameterizations since $\ud \tilde{t} = {\partial \tilde{t} \over \partial x^\mu} \ud x^{\mu} = \ud t$.

\item The spacetime metric $g_{\mu \nu}$ that allows us to construct scalar quantities. 
    
\end{enumerate}

Contracting $\ud t$ with arbitrary tensors allows us to form invariant scalars. For example, contracting $\ud t$ with the inverse metric gives its norm, i.e.~$g^{00}$, which behaves as a scalar under spatial diffeomorphisms. Therefore, we can equivalently use the unit vector $n_{\mu} = - (-g^{00})^{-1/2} \delta^0_{\mu}$ as a building block. This vector can be used to define the projector tensor on  spacelike hypersurfaces orthogonal to $n_{\mu}$:
\be
P_{\mu \nu} \equiv g_{\mu \nu} + n_{\mu} n_{\nu} \, .
\ee
%
When doing so, the inverse metric component $g^{00}$ becomes equal to $-\mathcal{N}^{-2}$, where $\mathcal{N}$ is the lapse of the Arnowitt-Deser-Misner (ADM) decomposition~\cite{Arnowitt:1962hi}. The EOM of the (closed) EFT of inflation is
\be
\mathcal{E}_{\mu \nu} = G_{\mu \nu} + (3H^2 + 2\dot{H} ) g_{\mu \nu} + 2 \dot{H} \delta^0_{\mu} \delta^0_{\nu} - \dot{H}\delta g^{00} g_{\mu \nu} + \text{model dependent}\, ,
\ee
where the ``model dependent" part is derived from  varying terms involving higher powers of $\delta g^{00}$ and other contractions of the extrinsic curvature $K_{\mu \nu} \equiv P^{\alpha}_\mu \nabla_{\alpha} n_{\nu}$. To quadratic order, there are three possibilities: $\delta K_{\mu \nu} \delta K^{\mu \nu}$, $\delta g^{00} \delta K$, and $\delta K^2$ with $K \equiv K^\mu{}_\mu$. Because these equations are derived from a single action of a 3-dimensional diffeomorphism-invariant theory, they automatically satisfy the reduced identity
\be \label{eq:divergence_free}
\nabla^{\mu} \mathcal{E}_{\mu i} = 0 \, ,
\ee
i.e.~the spatial part of the divergence equation, since time diffeomorphisms are broken.\footnote{Starting from the action, the vector parameterizing allowed diffeomorphisms is expressed in the unitary gauge  as $\xi^{\mu}=(0,\xi^i)$. From this we find the off-shell identity of EOM as the spatial part of the divergence equation, i.e.~\eqref{eq:divergence_free}. In a different gauge, the diffeomorphism vector splits to $\xi^{\mu}=\xi_{\rm n} n^{\mu}+P^{\mu}_i \xi^i$, with $\xi_{\rm n} \equiv \xi^\mu n_\mu$, and the divergence equation becomes $P^{\mu}_i\nabla^{\nu}\mathcal{E}_{\mu \nu}=0$, which includes the unitary gauge as a special case.}

Although we build the action using terms that are invariant under spatial diffeomorphisms, in practice, we do not include all of them. Allowing for example time derivative of the lapse would lead to an extra scalar degree of freedom in addition to the two tensors and the one scalar degree of freedom in standard inflationary theory. The three propagating degrees of freedom of the EFT of inflation are based on the existence of the Hamiltonian and momentum constraints, which prohibit terms such as $\dot{\mathcal{N}}$ in the constraint equations. To ensure that such term will not be present, contractions will involve objects projected on the 3-dimensional hypersurface and powers of $\delta g^{00}$.

\section{Open gravitational dynamics} 
\label{sec:consistent_eft}

To construct an open EFT of inflation, as a first step, we write the action as an expansion in the advanced metric, which is treated as a fluctuating matter field. In general, after integrating out environmental degrees of freedom, the effective dynamics of the remaining fields, which in unitary gauge are encoded in the metric, is described by a non-local influence functional \cite{Feynman:1963fq,Caldeira:1983uj}. Here we focus on the regime in which the environmental degrees have relaxation times much smaller than those of the system of interest, which for inflation is set by the Hubble parameter, that is \(H t_{\rm rel}\ll 1\).  
Under these assumptions, the relevant kernels admit a long-wavelength expansion, so that dissipative and stochastic effects can be represented by local operators. 

Up to quadratic order in perturbations the action takes the form
\be 
\label{eq:eff_action} 
S = {1\over 2} \int \ud^4x \sqrt{-g}  \left(  E_{\mu \nu} g_{\rm a}^{\mu \nu} +  g_{\rm a}^{\mu \nu} \mathcal{N}_{\mu \nu \kappa \lambda} g_{\rm a}^{\kappa \lambda} + \cdots \right) \, ,
\ee 
where $\mathcal{\mathcal{N}_{\mu \nu \kappa \lambda}}$ is the noise kernel and $E_{\mu\nu} \equiv 2(-g)^{-1/2}\, \delta S/\delta g_{\rm a}^{\mu\nu}\big|_{g_{\rm a}=0}$, so that $E_{\mu\nu}=0$ is the linear deterministic EOM. Both the deterministic sector and the noise kernel are constructed using the building blocks of the standard EFT of inflation to ensure invariance of the theory under (physical) time-dependent spatial reparameterizations. The EOM, in particular, is schematically 
\be
E_{\mu \nu} = \mathcal{E}_{\mu \nu} + \text{open terms} \, ,
\ee 
where ``open'' refers to terms appearing in the influence functional that arises after integrating out unobserved degrees of freedom. Drawing the parallel from systems with global symmetries where open terms explicitly break the advanced symmetry \cite{Sieberer:2015hba,Hongo:2018ant,Delacretaz:2021qqu}, we consider open terms that have non-zero covariant divergence, i.e.~they break advanced diffeomorphisms. This guarantees that such terms belong to the influence functional, since they do not respect the advanced diffeomorphism symmetry of the closed action, however, we should note that not every possible term in the influence functional necessarily breaks that symmetry.  

A complete local theory would require listing all possible open terms for which the EOM remains well-posed, namely, it is solvable with a unique solution that depends continuously on the parameters of the problem. This is the most challenging part of the EFT construction because in gauge theories or gravity, unless we choose open terms such that a (possibly new deformed) identity between the EOMs exists, we may encounter inconsistencies. These are manifested as an overconstrained system in terms of gauge-invariant combinations that represent the observables of our system.

\subsection{Why generic terms overconstrain the dynamics}

To understand why arbitrary terms overconstrain the EOM, we examine an open theory with a scalar field plus gravity. The incompatibility of dissipation for arbitrary open deformations of the scalar field EOM has also been shown in a closely related work using the St\"uckelberg trick \cite{Lau:2024mqm}. Here, we investigate this incompatibility by examining when the EOMs form a solvable system of equations. Assume the EOMs of the closed theory 
\be
\begin{aligned}
\mathcal{E}_{\mu \nu} & \equiv G_{\mu \nu} - T_{\mu \nu} = 0 \, , 
\\ 
\mathcal{E}_\phi & = 0 \, ,
\end{aligned}
\ee
where $\mathcal{E}_{\mu \nu}$ and $\mathcal{E}_\phi $ are the Euler-Lagrange expressions for the metric and scalar field, respectively. 
For a FLRW background performing the scalar-vector-tensor decomposition, we then find 3 scalar, 2 vector, and 2 tensor gauge-invariant perturbations in the EOMs (and of course at the level of the action). We can extend the minimal model beyond Einstein gravity and remain in the closed theory by adding new terms, $\tilde{\mathcal{E}}_{\mu \nu} \equiv \mathcal{E}_{\mu \nu} + \Delta \mathcal{E}_{\mu \nu}$, in both equations and as long as the divergence of the metric part $\nabla^{\mu} \mathcal{\tilde{E}}_{\mu \nu}$ is zero off-shell (i.e.~without using the metric that solves its EOM) but for $\phi$ that satisfies its EOM $\mathcal{\tilde{E}}_\phi =0$; the theory remains consistent. This holds because differential identities between the EOMs guarantee that the number of linearly independent combinations of the system variables matches the number of independent equations, thus yielding a consistent system.

In open theory, the EOMs can, in principle, be modified to leading order as follows:
\be \label{eq:open_scalar_gravity}
\begin{aligned}
E_{\mu \nu} &\equiv \mathcal{E}_{\mu \nu} + \gamma_1 R g_{\mu \nu} + \gamma_2 T(\phi) g_{\mu \nu} + \gamma_3 \nabla_{\mu} \nabla_{\nu} \phi + \dots = 0 \, , \\
E_\phi &\equiv \mathcal{E}_\phi + \gamma_{\phi} \sqrt{-\partial_{\mu}\phi \partial^{\mu}\phi} +\dots = 0 \, .
\end{aligned}
\ee
Regarding the scalar field equation, the second term is reduced to the familiar form $\gamma_\phi n^{\mu}\partial_{\mu}\phi$
when choosing as normal vector the field velocity, which for FLRW background is timelike to zeroth order:
\be
n_{\mu} = -{\partial_{\mu} \phi \over \sqrt{-\partial_{\nu}\phi \partial^{\nu}\phi}} \, .
\ee
Solving the two sets of equations simultaneously in \eqref{eq:open_scalar_gravity} is not possible for an arbitrary open extension. More specifically, the four scalar EOMs for the metric part plus the EOM for the scalar field form a system of 5 equations for 5 unknowns, which is in principle solvable. However, because we only considered covariant terms, we should be able to perform coordinate transformations and fix some of the metric components. For this to be possible, the 5 field perturbations (metric and scalar) should appear in 3 linearly independent combinations. Thus, unless the 5 equations are related, the system of equations will be overconstrained,\footnote{Ref.~\cite{Salcedo:2025ezu} used a similar argument to explain why a deformed advanced symmetry necessarily appears in an open extension of Maxwell theory. While this holds true for all terms allowed by physical symmetries in that case, gravity is qualitatively different as we demonstrate in this work.} and crucially, this limits the form of available open terms. We illustrate this with a concrete example in Section~\ref{subsec:gammak} using the extrinsic curvature as our open term.
The same holds if the field fluctuations are set to zero, or equivalently when working in the unitary gauge, or if a St\"uckelberg field is introduced to restore time diffeomorphisms.

\subsection{Revisiting the extrinsic curvature term and the decoupling limit} 
\label{subsec:gammak}

Back to the EFT of inflation, when open terms are introduced, the coefficients of the universal part of the action will change to account for any non-zero background contribution from the extra terms. We consider the universal part of the action plus an additional open term:
\be 
\label{eq:eom_open}
E_{\mu \nu } = G_{\mu \nu} +  \bar{c}_{\delta} \delta^0_\mu \delta^0_\nu + \bar c_{\rm g} g_{\mu \nu} - {1\over 2} \bar{c}_{\delta} \delta g^{00} \bar{g}_{\mu \nu} + \Delta_{\mu \nu} \, ,
\ee
where $\nabla^\mu \Delta_{\mu i} \neq 0$. Imposing FLRW as our background results in the following two conditions:
\begin{align}
 3H^2 + \Bar{c}_{\delta} - \Bar{c}_{\rm g} + \bar{\Delta}_{00}&=0 \, , \\
-a^2(3H^2 + 2\dot{H})\delta_{ij} + \Bar{c}_{\rm g} a^2 \delta_{ij} + \bar{\Delta}_{ij} &= 0 \, ,
\end{align}
which fix the coefficients $\Bar{c}_{\delta}$ and $\Bar{c}_{\rm g}$.

\textbf{\textit{Extrinsic curvature}}:
One of the leading open terms that affects tensor perturbations is the extrinsic curvature $\Delta_{\mu \nu}=\Gamma K_{\mu \nu}$ \cite{LopezNacir:2011kk,Lau:2024mqm}; in Appendix~\ref{app:vectors} we explicitly show how it affects scalar and vector perturbations. This term is absent from the EOMs of the standard EFT of inflation because it can only be obtained  from the term $K_{ij} g^{ij}$ which is a total derivative. 
With this choice, the coefficients of the universal part are fixed to
\be
\begin{aligned}
\Bar{c}_{\delta} & = 2 \dot{H}  - H \Gamma \, , 
\\ 
\Bar{c}_{\rm g} & = 3H^2 + 2\dot{H} - H \Gamma  \, .
\end{aligned}
\ee
Next, we focus on scalar perturbations (see e.g.~\cite{Mukhanov:1990me}); we introduce scalar perturbations in the metric, ignoring vector and tensor perturbations since they decouple to linear order:
\be \label{eq:adm_metric}
\ud s^2 = -(1 + 2 A)\ud t^2 + 2 a \partial_i B \ud t\ud x^i 
+ a^2 \left[ (1 - 2 \psi) \delta_{ij} +  2\partial_{i} \partial_{j} \chi  \right] \ud x^i \ud x^j \, ,
\ee
and from \eqref{eq:eom_open} find the following linear system for scalar perturbations:
\begin{align} \label{eq:gmna}
&E_{00} =  {2\over a^2} \nabla^2 \psi + {2 H\over a^2} \nabla^2 \sigma -  6 H \left(\dot{\psi} + H A \right) - 2 \dot{H}A  + \Gamma H A  \, , \\
\label{eq:e0i}
&E_{0i} = 2 \partial_i (\dot{\psi} + H A) \, ,\\ \label{eq:eij}
&\begin{aligned}
E_{ij} &= 2 a^2 \left[\left(3 H - {1\over 2} \Gamma \right) \left(\dot{\psi} + H A \right) +  \partial_t\left(\dot{\psi} +  H A \right)\right]\delta_{ij} 
+ \mathcal{D}_{ij}\left(\psi - A  + H \sigma + \dot{\sigma} \right)
\\ 
& ~~~ + a^2 H \Gamma A \delta_{ij} + \Gamma \partial_i \partial_j \sigma \, ,    
\end{aligned}
\end{align}
where $\sigma$ is the shear perturbation defined by
\begin{equation}
\sigma \equiv a\left(a\dot{\chi} - B\right) \, ,
\end{equation}
and in the $E_{ij}$ equation we have introduced the operator 
\be
\mathcal{D}_{ij} \equiv \partial_i \partial_j - \delta_{ij} \nabla^2 \, ,
\ee
with the important property $\partial^i \mathcal{D}_{ij} = 0$.
Notice that among the scalar perturbations, $B$ and $\chi$ appear exclusively through the shear. Extracting the trace and traceless parts from \eqref{eq:eij} gives
\begin{align}
-2 \nabla^2\left(\psi - A  + H \sigma + \dot{\sigma} \right) + \Gamma \nabla^2 \sigma + 3 a^2H\Gamma A = 0 \, ,
\\ 
\label{eq:traceless_curv}
\left( \psi - A  + H \sigma + \dot{\sigma} \right) + \Gamma  \sigma = 0 \, ,
\end{align}
where we assumed appropriate boundary conditions to invert the Laplacian operators and used \eqref{eq:e0i} to set $\dot\psi+HA=0$. The previous two equations impose $\nabla^2 \sigma = - a^2 HA $, substituting which into \eqref{eq:gmna} and \eqref{eq:traceless_curv} gives two incompatible equations:
\begin{align}
{1\over a^2} \nabla^2 \psi + \dot{\psi}\left( H +  {\dot{H} \over H} -{1\over 2} \Gamma \right) = 0 \, , 
\\
\nabla^2 \left( \psi + {\dot{\psi} \over H} \right) +  a^2\dot{\psi} \left(H+\Gamma\right) +  \partial_t\left(a^2 {\dot{\psi} } \right) = 0 \, . 
\end{align}
This system of equations is satisfied only for the trivial solution $\psi =0$ and the theory fails to reproduce the expected scalar degree of freedom.


\textbf{\textit{The decoupling regime}}:  Potential inconsistencies introduced by certain open-system operators may be overlooked when working in the decoupling limit. This is reminiscent of theories with higher derivatives, where the existence of a healthy decoupling limit is not sufficient to demonstrate stability of the full theory \cite{Gumrukcuoglu:2013nza,deRham:2014naa}. In this regime the constraint structure of the full gravitational theory is no longer manifest because metric fluctuations have been ignored. To demonstrate this, we can consider an alternative simple open term in \eqref{eq:eom_open}:
\be
\Delta_{\mu \nu} = \Gamma \delta g^{00} \left(\bar{g}_{\mu \nu} + \delta_\mu^0 \delta_\nu^0 \right)
\ee
with $\Gamma \neq 0$.\footnote{Note that in the closed theory varying the term $F[\delta g^{00};t]$ in the action produces the following EOM:
\begin{equation*}
{\partial F \over \partial \delta g^{00}} \delta_\mu^0 \delta_\nu^0 - {1\over 2} F g_{\mu \nu} \, .
\end{equation*}
At linear order, any term involving $\delta g^{00} g_{\mu \nu}$ whose coefficient does not respect the  above structure is automatically open.}
This simple model was examined in \cite{Salcedo:2025ezu} and showed that it reproduces the basic structure of the open EFT of inflation written in terms of the Goldstone boson in the decoupling limit \cite{Salcedo:2024smn}. Here we examine the same model in the unitary gauge, keeping all gravitational degrees, and find the following EOMs for scalar perturbations from \eqref{eq:eom_open}:
\begin{align}
&E_{00} =  {2\over a^2} \nabla^2 \psi + {2 H\over a^2} \nabla^2 \sigma -  6 H \left(\dot{\psi} + H A \right) - 2 \dot{H}A \, , \\
&E_{0i} = 2 \partial_i (\dot{\psi} + H A) \, ,\\
&\begin{aligned}
E_{ij} &= 2 a^2 \left[ \Gamma A + 3 H \left(\dot{\psi} + H A \right) +  \partial_t\left(\dot{\psi} +  H A \right)\right]\delta_{ij} 
+ \mathcal{D}_{ij}\left(\psi - A  + H \sigma + \dot{\sigma} \right)
\, .  
\end{aligned}
\end{align}
Combining the trace and traceless parts of the $ij$ equation we find $A=0$, which subsequently forces $\psi = \sigma =0$ as the only solution. Note that only the term $\delta g^{00} g_{\mu \nu}$ is responsible for the absence of a propagating mode, while the term proportional to $\delta g^{00} \delta_\mu^0 \delta^0_\nu$ affects only $E_{00}$ which is not part of the identity equation at linear order in perturbations. In the $\pi$-language of \cite{Salcedo:2025ezu} this inconsistency is invisible because metric fluctuations are neglected. Our full analysis shows that the open term $\delta g^{00} g_{\mu\nu}$ actually removes the scalar mode once gravity is reinstated.

The mechanism behind the failure to produce a consistent system in both cases can be understood by examining the $0i$ and $ij$ components of the EOM. In the closed theory these take the schematic form (see Appendix~\ref{app:scalars})
\begin{align} 
E_{0i} &= 2\partial_i Z \, , \\ \label{eq:structure}
E_{ij} &= \mathcal{D}_{ij} C + a^2 F(Z,\dot{Z}) \delta_{ij} \, ,
\end{align}
where $C$ is a function of the perturbation variables and $F$ is a linear function of $Z \equiv \dot{\psi} + H A$ and its derivative. Taking the spatial divergence of \eqref{eq:structure} removes the $C$-dependent part, leaving an expression that can be written entirely in terms of $E_{0i}$, thus generating an identity. The two open terms discussed here do not preserve this structural relation and, therefore, the two equations cannot be related, leading to an overconstrained system.

Before presenting our proposal for a gravitational EFT in the next section, we clarify how it differs from previous constructions. Ref.~\cite{Lau:2024mqm} proposed accounting for dissipation in the presence of dynamical gravity by embedding the dissipative system into a larger system modelled through hydrodynamic EFT, an approach that is naturally suited to describe memory effects and environmental fluctuations. In that construction the environmental variables are kept explicitly: once a constitutive relation and an equation of state are specified, the scalar, gravitational and hydrodynamic sectors form an enlarged closed set of equations. Here, instead, the environment is integrated out and we seek a reduced, local EFT for the metric sector alone, whose consistency is ensured by a deformed identity between the EOMs rather than by additional dynamical variables. The two approaches are therefore complementary, with the present one restricted to the local (Markovian) regime described above. By contrast, Ref.~\cite{Salcedo:2025ezu} proposed to construct an open EFT by writing all terms compatible with the physical symmetries, without requiring the existence of deformed identities between EOMs. When the latter symmetry requirement is imposed, the otherwise extensive set of operators allowed by the physical symmetries becomes severely restricted. 

\section{An open EFT of inflation with gravity} 
\label{sec:open_eft}

In the previous section, we argued that a system of equations in the open theory is consistent given that the EOMs satisfy an off-shell identity, which requires careful selection of open terms. Because such terms are introduced as continuous deformations of the closed theory, the desired identity should be a deformation of \eqref{eq:divergence_free}, and should be built from the same building blocks as our theory. Thus, it should take the form
\be \label{eq:identity_projector}
P^\nu_i\nabla^\mu E_{\mu \nu} = P^\nu_i \left[ \alpha E_{\mu \nu}n^\mu + \beta \left(E_{\kappa \lambda} g^{\kappa \lambda}\right)_{,\nu} + \delta n^\kappa \nabla_{\kappa} (E_{\mu \nu} n^\mu) + \cdots \right] \, .
\ee
Notice that in the right hand side there is no term proportional to $n_{\nu}$, because it vanishes once we act with the projector. Some of the open terms that allow for non-perturbative identities have been presented in \cite{Christodoulidis:2025ymc}, and they are related to known identities of the ADM formalism or they involve algebraic contractions of the EOMs along the normal vector or orthogonal subspace. In this section, we are interested in the minimal open extension of the standard EFT of inflation that modifies the EOMs for both scalar and tensor perturbations and thus only consider terms that can yield an identity involving the $\alpha$ term in \eqref{eq:identity_projector}.\footnote{Terms involving the trace or other projections of the EOMs are trivial in the sense that they do not affect the equations of propagating modes \cite{Christodoulidis:2025ymc}.}

\subsection{Constructing an EFT with a propagating scalar}

To obtain a minimal open EFT model we consider the following extrinsic curvature combination:
\be \label{eq:trace_adjusted}
\Delta_{\mu \nu}  =  {1 \over \mathcal{N}} \left( K_{\mu \nu} - K P_{\mu \nu} \right)  \, ,
\ee
with $\cal{N}$ denoting the lapse function, which satisfies the exact identity \cite{Christodoulidis:2025ymc}
\be \label{eq:divergence_extr_curv}
P^\nu_i \nabla^\mu \left( {K_{\mu \nu} - K P_{\mu \nu} \over \mathcal{N}} \right) = {1 \over \mathcal{N}} D^j \left( K_{ij} - K P_{ij} \right) = {1 \over \mathcal{N}}  P^\nu_i G_{\mu \nu} n^\mu \, .
\ee
Using the relation between the lapse and the metric coefficient with upper indices, $\mathcal{N}^{-2} = -g^{00}$, our open extension of the EFT of inflation takes the form 
\be  \label{eq:EOM_extr}
\begin{aligned}
E_{\mu \nu } =~&  
\underbrace{ G_{\mu \nu} 
+  \left(\bar{c}_\delta +M^2 \delta g^{00} \right) \delta^0_\mu \delta^0_\nu 
+ \left[ \bar{c}_{\rm g} - {1\over 2} \bar{c}_\delta  \delta g^{00} - {1\over 4} M^2 \left( \delta g^{00} \right)^2 \right] 
g_{\mu \nu} }_{
\text{closed part}
}
\\
& + \underbrace{ \Gamma ( K_{\mu \nu} - K P_{\mu \nu})\sqrt{-g^{00}} }_{
\text{open part}
} \, .
\end{aligned}
\ee
In the latter expression we also allowed for an arbitrary $M$ that is related to the speed of sound of the closed theory and, therefore has zero divergence (projected on the hypersurface) identically.\footnote{Here we work in units with $\mpl=1$ and $M$ denotes $M \equiv \sqrt{2}M_2^2(t)/\mpl$, where $M_2(t)$ is the coupling of the dimension 4 operator of the closed EFT of inflation, $1/2 M_2^4(t) (\delta g^{00})^2$ \cite{Cheung:2007st}.} Before fixing the background let us prove that this minimal extension admits off-shell identities between the EOMs. Projecting \eqref{eq:EOM_extr} along the normal vector and then along the hypersurface gives
\be
P^\nu_i E_{\mu \nu} n^\mu = P^\nu_i G_{\mu \nu} n^\mu \, ,
\ee
which is the right hand side of the divergence of the open term \eqref{eq:divergence_extr_curv}. 
Therefore, the EOM of our system satisfies the \textit{deformed diffeomorphism identity}
\be \label{eq:deformed_identity}
P^\nu_i \nabla^{\mu} E_{\mu \nu} = {\Gamma  \over \mathcal{N} } P^\nu_i  E_{\mu \nu} n^{\mu} \, ,
\ee
valid at all orders in perturbation theory. Note that this is the exact analogue of the deformed identity in the open Maxwell theory $\partial_\mu \mathcal{E}^\mu_{\rm M} = \Gamma u_\mu \mathcal{E}^\mu_{\rm M}$, with $\mathcal{E}^\mu_{\rm M}$ denoting its EOM \cite{Salcedo:2024nex}.

Because we introduced the extrinsic curvature, imposing FLRW as our background results in the following two conditions, which fix $\bar{c}_{\delta}$ and $\bar{c}_{\rm g}$:
\begin{align}
 3H^2 + \Bar{c}_{\delta} - \Bar{c}_{\rm g} &=0 \, , \\
-a^2(3H^2 + 2\dot{H})\delta_{ij} + \Bar{c}_{\rm g} a^2 \delta_{ij} -2  a^2 H \Gamma \delta_{ij} &= 0 \, .
\end{align}
The two coefficients are found as
\be
\begin{aligned}
\Bar{c}_{\delta} & = 2 \dot{H} +2 H \Gamma \, , 
\\ \Bar{c}_{\rm g} & = 3H^2 + 2\dot{H} +2 H \Gamma  \, ,
\end{aligned}
\ee
and $\Gamma$ is left as a free parameter. 

We continue by expanding the EOMs to linear order in metric perturbations. 
We adopt 
\eqref{eq:adm_metric} and define a 3-dimensional operator that extracts the traceless part of a spatial tensor:
\be
\mathcal{T}_{ij} \equiv \nabla^{-2}\left(\partial_i\partial_j - {1\over 3}\nabla^{2}\delta_{ij} \right) \, ,
\ee
where by definition $\delta^{ij}\mathcal{T}_{ij}=0$. The part involving the advanced fields in the EOM, 
\be
\Xi_{\mu \nu} \equiv - \left(\mathcal{N}_{\mu \nu \kappa \lambda} + \mathcal{N}_{\mu \nu \kappa \lambda}^\dag \right) g_{\rm a}^{\kappa \lambda} \, ,
\ee
can be decomposed similarly. Expanding in metric perturbations and solving the constraints we find the EOM of the scalar degree of freedom (all calculations are presented in Appendix~\ref{app:scalars}):
\be \label{eq:curv_pertur}
\ddot{\psi} +  \left\{ ( 3 + \gamma  )H + \partial_t \left[\log \left( {2\epsilon + 2\gamma \over  c_{\rm s}^2 } \right) \right] \right\} \dot{\psi} - c_{\rm s}^2 {\nabla^2 \psi \over a^2}   = \Xi_{\psi} \, ,
\ee
with the definitions
\begin{align}
\gamma & \equiv {\Gamma \over H} \, , 
\\  \label{eq:sound_speed}
c_{\rm s}^2 & \equiv {\epsilon + \gamma \over \epsilon +M^2/H^2 -\gamma} \, ,
\\
\Xi_{\psi} & \equiv 
a^{-2} \bigg( 3 - \frac{\epsilon+\gamma}{c_{\rm s}^2} \bigg)^{-1}
\Bigg\{
{3\over 2} \mathcal{T}^{ij}\Xi_{ij} + {\partial^i\Xi_{0i} \over 2H}  
\nonumber\\
& \hspace{10em}
- \bigg[ {\ud \over \ud t} + H(1+\gamma) \bigg]
\left[ {a^2 \over 2H} \Xi_{00} + {a^2 \over 2} \bigg( 3 - \frac{\epsilon+\gamma}{c_{\rm s}^2} \bigg) \nabla^{-2}\partial^i\Xi_{0i} \right] \Bigg\}
\end{align}
This is our main result. In the closed EFT limit, both $\Xi_{\psi}$ and $\Gamma$ are set to zero and the previous reduces to the standard EOM for the curvature perturbation $\psi = -\mathcal{R}$ in the EFT of inflation.

\textbf{\textit{The decoupling regime}}: It is also useful to examine the decoupling limit of the present open theory. To restore broken time diffeomorphisms, we introduce two St\"uckelberg fields, $\pi$ and $\pi_{\rm a}$ and then proceed in two steps. First, starting from the SK action \eqref{eq:eff_action}, we perform a deformed advanced diffeomorphism,
\begin{equation}
\delta g_{\rm a}^{\mu \nu} = - 2\nabla^{(\mu} \xi_{\rm a}^{\nu)} - 2 \Gamma n^{(\mu} \xi_{\rm a}^{\nu)} \, .
\end{equation}
Restricting to the deterministic sector, this transformation leaves at quadratic order the residual term
\begin{equation} 
\label{eq:SK_goldstone}
\delta S_{\rm a} = {1\over 2} \int \ud^4x \sqrt{-g} \, \xi^{0}_{\rm a} n^{\nu}\left( \nabla^{\mu} - \Gamma n^\mu \right) E_{\mu \nu} \, .
\end{equation}
Promoting $\xi_{\rm a}^0$ to a new field $\pi_{\rm a}$, we extend the deterministic part of the action to
\begin{equation}
S = {1\over 2} \int \ud^4x \sqrt{-g} \left[ E_{\mu \nu} g_{\rm a}^{\mu \nu} + \pi_{\rm a} n^{\nu}\left( \nabla^{\mu} - \Gamma n^\mu \right) E_{\mu \nu} \right]\, ,
\end{equation}
and assigning $\delta\pi_{\rm a}=-\xi_{\rm a}^0$, the action becomes invariant under 4D deformed advanced diffeomorphisms. Second, we restore physical time diffeomorphisms by the standard replacement $f(t)\to f(t+\pi)$, together with
\begin{equation}
n_\mu=-\mathcal{N} \delta_\mu^0 \;\to\; -\widetilde{\mathcal{N}} \Pi_\mu
\, ,
\end{equation}
where $\widetilde{\mathcal{N}}^{-2}\equiv -g^{\mu\nu}\Pi_\mu \Pi_\nu$ and $\Pi_\mu\equiv \delta_\mu^0+\partial_\mu\pi$.

The EOM for $\pi$ is obtained by varying the action with respect to $\pi_{\rm a}$. Expanding to quadratic order in scalar perturbations, the action takes schematically the form
\begin{equation}
S[\pi,\pi_{\rm a},g_{\mu \nu},g_{\rm a}^{\mu \nu}] = S_\pi[\pi,\pi_{\rm a}]  + S_{\rm mix}[\pi,\pi_{\rm a},g_{\mu \nu},g_{\rm a}^{\mu \nu}] \, .
\end{equation}
Isolating the terms that depend only on $\pi$ and $\pi_{\rm a}$, we obtain the Goldstone sector:
\begin{equation}
S_\pi[\pi,\pi_{\rm a}] 
= 
\int \ud^4x \,a^3 \theta \pi_{\rm a}  \left\{ \ddot{\pi} +  \big[ (3+ \gamma)H + \partial_t(\log \theta) \big] \dot{\pi} - c_{\rm s}^2 {\nabla^2 \pi \over a^2} + 3 {1\over \theta} \left[\dot{H}^2 - H^2 \partial_t{(\gamma H) } \right] \pi \right\}
\, ,
\end{equation}
where
\begin{equation} \label{eq:theta_def}
\theta \equiv M^2 - H^2 \gamma - \dot{H} 
= H^2 \frac{\epsilon+\gamma}{c_{\rm s}^2}
\,, 
\end{equation}
and the sound speed is the same as in~\eqref{eq:sound_speed}. In the limit of vanishing dissipation, this reproduces the standard EOM for the Goldstone boson $\pi$ in the decoupling regime (see   e.g.~\cite{Green:2024hbw}).

The mixing part is
\begin{equation}
S_{\rm mix} = \int \ud^4x \, a^3 \pi_{\rm a} \left\{ -{ H^2 \over a} {\ud \over \ud t} \left({\nabla^2 B \over H}\right) -  A \left[ \dot{\theta} + (3+\gamma)H \theta - 3 H\dot{H} \right] - 3 H^2 \dot{A} \right\} \, .
\end{equation}
Estimating the regime in which decoupling is valid requires solving the constraint equations to linear order:
\begin{align}
\label{eq:constraint_A}
A & = \epsilon H \pi \, ,
\\
\label{eq:constraint_B}
\nabla^2 B & = -{a \theta \over H^2} \partial_t(H\pi)
\, ,
\end{align}
where $A$ and $B$ are the scalar metric perturbations in \eqref{eq:adm_metric} and we work in the flat gauge, where $\chi=\psi=0$. Note that the momentum constraint \eqref{eq:0i-excurv} is not deformed by the open terms, so \eqref{eq:constraint_A} is unchanged with respect to the closed theory, while the open parameters enter the energy constraint \eqref{eq:00-excurv}, and hence \eqref{eq:constraint_B}, only through the overall factor $\theta$ defined in \eqref{eq:theta_def}; for $\Gamma = M = 0$ one has $\theta = -\dot{H}$ and \eqref{eq:constraint_B} reproduces the standard single-field result. Both expressions are therefore exact at linear order and no assumption on the size of $\Gamma$ and $M$ is needed at this stage. Substituting the previous into $S_{\rm mix}$ one can compare the induced mixing terms with the intrinsic $\pi$ dynamics. This is meant as an illustration of how the mixing terms can be estimated rather than as a full derivation of the decoupling conditions, whose detailed analysis we leave for future work.

\subsection{The scalar power spectrum}

To calculate the power spectrum we need to derive the general solution of \eqref{eq:curv_pertur}, which can be written as
\be
\mathcal{L} \psi = \Xi_{\psi} \, ,
\ee
with
\be 
\mathcal{L} \equiv \partial_t^2 +  \left\{ ( 3 + \gamma  )H + \partial_t \left[\log \left( {2\epsilon + 2\gamma \over  c_{\rm s}^2 } \right) \right] \right\} \partial_t - c_{\rm s}^2 {\nabla^2  \over a^2}  \, .
\ee
The solution is found via the Green's method as the sum of the homogeneous and particular solutions:
\be
\psi = \psi_{\rm hom} + \int_{\tau_0}^\tau \ud \tau '\, G(\tau,\tau ') \Xi_{\psi}(\tau ') \, ,
\ee
each satisfying 
\be
\begin{aligned}
\mathcal{L} \psi_{\rm hom} & = 0 \, , 
\\   
\mathcal{L} G(\tau,\tau') & = \delta (\tau-\tau')\, .
\end{aligned}
\ee
Assuming slow-roll conditions on the various parameters in the differential equation for $\psi$, then up to slow-roll corrections we find the general solution of the homogeneous system as a superposition of Hankel functions of the first and second kind $\psi_{\rm hom}^{\pm} = (-\tau)^{\nu }\mathrm{H}^{(1,2)}_{\nu }(-c_{\rm s}k\tau)$ where
\be
\nu \equiv {1\over 2} ( 3 + \gamma )  
\, .
\ee
The homogeneous solution is \eqref{eq:homogeneous_general} of Appendix~\ref{app:homogeneous}. The Green's function can be found using the Wronskian: 
\be
G(\tau,\tau') = \Theta(\tau - \tau')\, a^2(\tau') \left( {\tau \over \tau'} \right)^{\nu} {\mathrm{H}^{(2)}_{\nu }(x)\mathrm{H}^{(1)}_{\nu }(x') - \mathrm{H}^{(1)}_{\nu }(x)\mathrm{H}^{(2)}_{\nu }(x')  \over \mathrm{H}^{(1)}_{\nu }(x') \, \partial_{\tau'}\mathrm{H}^{(2)}_{\nu }(x') -\mathrm{H}^{(2)}_{\nu }(x') \, \partial_{\tau'}\mathrm{H}^{(1)}_{\nu }(x') } \, ,
\ee
with $x \equiv -c_{\rm s}k\tau$ and $x' \equiv -c_{\rm s}k\tau'$.
Of course, one could equally work directly with the SK action and calculate the power spectrum from the Keldysh propagator (see e.g.~\cite{Salcedo:2024smn}).

In line with the standard open-system treatment of inflationary perturbations (see e.g.~\cite{Calzetta:1995ys,Boyanovsky:2015tba}) we assume a factorized initial density matrix, so that we can assign Bunch-Davies initial condition for the curvature perturbation. The normalized power spectrum is given as the sum of two contributions, one related to Bunch-Davies initial condition and one sourced by noise:\footnote{Generic initial states would induce cross-correlations between noise and $\psi$, which would be suppressed by the dissipation parameter $\gamma$.} 
\begin{equation}
\mathcal{P}_{\mathcal{R}} 
=
\left\{ 1 + \left[ \frac{8(c_{\rm s}-\tilde{c}_{\rm s}) - \gamma \tilde{c}_{\rm s}(6+\gamma)}{8\tilde{c}_{\rm s}\gamma\omega_0} \right]^2 \right\}
\left( \frac{2}{\gamma\omega_0} \right)^\gamma \left[ \frac{2}{\sqrt{\pi}} \Gamma \left( \frac{3+\gamma}{2} \right) \right]^2
\frac{\tilde{c}_{\rm s}}{c_{\rm s}}
\mathcal{P}_0
+ \sum_i \mathcal{P}^i_{\rm noise} \, ,
\end{equation}
where $\omega_0 \equiv -c_{\rm s}k\tau_0/\gamma$, and $\mathcal{P}_0$ is the standard single-field power spectrum:
\be
\mathcal{P}_0 \equiv {H^2 \over 8 \pi^2 \epsilon c_{\rm s}} \, .
\ee
Here, $\tilde{c}_{\rm s}$ and $c_{\rm s}$ denote respectively the sound speeds before and after $\tau_0$, and $i$ runs from 1 to 3, which is the maximum number of independent noise components in the scalar sector.\footnote{Recall that the deformed identity \eqref{eq:deformed_identity} forces the noise components (through the EOMs) to satisfy the relation
\begin{equation*}
P^\nu_i \left( \nabla^\mu - {\Gamma \over \mathcal{N}} n^\mu \right) \Xi_{\mu \nu} = 0 \, ,
\end{equation*}
and therefore their independent scalar components are reduced by one.} The noise-induced part of the power spectrum depends on the specific form of the noise kernel $\mathcal{N}_{\mu\nu\kappa\lambda}$, so its explicit form is model dependent (see e.g.~\cite{LopezNacir:2011kk,Creminelli:2023aly,Salcedo:2024smn} for explicit expressions). The total power spectrum is dominated by noise for large dissipation, as the homogeneous solution is exponentially suppressed. For small dissipation, depending on the magnitude of the noise contribution, both parts can be equally important.

\section{Summary and discussion} 
\label{sec:summary}

In this work, we have shown that in the SK formulation of the EFT of inflation, most open operators allowed by the physical symmetries are incompatible with gravitational dynamics once the full constraint structure is taken into account. Generic open deformations overconstrain the scalar sector as illustrated by the two examples in Section~\ref{subsec:gammak}. In contrast, we identified a minimal ``trace-adjusted'' extrinsic-curvature combination \eqref{eq:trace_adjusted} that generates a non-trivial deformed diffeomorphism identity and thus guarantees the correct number of degrees of freedom. This operator defines a dynamically consistent open extension of the EFT of inflation beyond the decoupling limit, modifies all three sectors, and yields a Langevin-type equation for the curvature perturbation with dissipation and noise. 


A natural next step is to move beyond this minimal construction and systematically classify all open operators of the local EFT of inflation. We have also observed that for the purely spatial term 
\be
S_{\mu \nu}  \equiv P^\alpha_\mu  P^\beta_\nu R_{\alpha \kappa \beta \lambda} n^\kappa n^\lambda \, ,
\ee
its trace adjusted combination
\be
S_{\mu \nu} - S P_{\mu \nu} - H^2 \, \delta g^{00} P_{\mu \nu} \, ,
\ee
satisfies, at linear order, a deformed identity of the form \eqref{eq:identity_projector}. More specifically, adding this term with coefficient $\Gamma$ to the EOM generates the identity \eqref{eq:identity_projector} with $\alpha=-3H\Gamma$, $\delta=-\Gamma$ and all other coefficients set to zero at linear order. Note that this term appears in the second Bianchi identity $\nabla_{[\mu}R_{\nu \kappa] \lambda \rho} = 0$ when it is contracted with $g^{\mu \lambda} n^\kappa n^\rho$. Other terms consistent at quadratic order are the parity-violating terms found in \cite{Salcedo:2025ezu}.
It would be interesting to investigate if all previous remain consistent  beyond linear level, thus providing us with additional non-trivial open terms.

We also expect that the consistent open EFT developed in this work will be useful for identifying possible nonlinear interactions involving dissipation and noise, and for computing their associated observable signatures. A primary target in this regard is the three-point correlation function, i.e.~the bispectrum. Certain ultraviolet-complete models predict a large scalar bispectrum, for example, in the equilateral configuration~\cite{Creminelli:2023aly}, while different shapes have been reported depending on the strength of dissipation~\cite{Salcedo:2024smn}. However, the full set of bispectrum shapes allowed within a consistent open EFT of inflation remains largely unexplored. Moreover, extensions of general relativity generically introduce additional interaction structures that can give rise to novel bispectral shapes, particularly mixed scalar–tensor correlations (see e.g.~\cite{Gong:2023kpe,Christodoulidis:2024ric}). It would therefore be especially interesting to understand how extended theories of gravity interact consistently within an open-system framework, and how such interactions may lead to distinctive bispectra that could be tested by future observations.

\acknowledgments
PC thanks Enrico Pajer and Thomas Colas for useful discussions.  
%
We are supported in part by the Basic Science Research Program through the National
Research Foundation of Korea (RS-2024-00336507).
JG is grateful to the Asia Pacific Center for Theoretical Physics for hospitality while this work was in progress.

\begin{appendix}
\section{Cosmological perturbation theory}

In this section, we provide analytical calculations for the scalar and vector modes. The metric is expanded in scalar, vector and tensor perturbations
\be \label{eq:adm_metric2}
\ud s^2 = -(1 + 2 A)\ud t^2 + 2 a (\partial_i B + v_i)\ud t\ud x^i + a^2 \left[ \delta_{ij}(1 - 2 \psi) +  2\partial_{i} \partial_{j} \chi +  2\partial_{(i} w_{j)} + \gamma_{ij} \right] \ud x^i \ud x^j \, ,
\ee
where vectors and tensors satisfy transverse or traceless conditions
\be
\partial^i v_i = \partial^i w_i = \partial^i \gamma_{ij} = 0 \, , \quad \delta^{ij} \gamma_{ij} = 0 \, .
\ee

\subsection{Scalars} \label{app:scalars}

The scalar part of the EOMs of the standard EFT of inflation to linear order is written entirely in terms of only three variables $A$, $\psi$ and $\sigma$:\footnote{For a maximally symmetric space ($\dot{H}=0$), the shear perturbation can be absorbed by defining the following two variables:
\begin{align*}
\Psi & \equiv \psi + H \sigma \, ,
\\ 
\Phi & \equiv A - \dot{\sigma} \, ,
\end{align*}
which, combined with EOMs, imply that there are no dynamical scalar perturbations.}
\begin{align}
& \mathcal{E}_{00} =  {2\over a^2} \nabla^2 \psi + {2 H\over a^2} \nabla^2 \sigma -  6 H \left(\dot{\psi} + H A \right) - 2 \dot{H}A\, , \\ 
& \mathcal{E}_{0i} = 2\partial_i \left(\dot{\psi} + H A \right) \, , \\
& \mathcal{E}_{ij} = 2 a^2 \left[3 H \left(\dot{\psi} + H A \right) + \partial_t\left(\dot{\psi} + H A \right)\right]\delta_{ij} 
+ \mathcal{D}_{ij}\left(\psi - A  +  H \sigma +  \dot{\sigma} \right) \, .
\end{align}

Let us pause and investigate the constraint structure and solutions of the previous set of equations. The latter satisfy the \textit{scalar identity}:
\be \label{eq:scalar_identities}
\partial^j \mathcal{E}_{ij} = a^2 \left( 3H \mathcal{E}_{0i} + \dot{\mathcal{E}}_{0i} \right) \, ,
\ee
which is the spatial part of the divergence equation
\be
\nabla^{\mu} \mathcal{E}_{\mu i} = 0 \, ,
\ee
for scalar modes; the zeroth component of the divergence equation is explicitly broken. We also derive the EOM for the dynamical degree of freedom, the curvature perturbation. The second equation yields $A = -\dot{\psi}/H$ which, substituting into the first, gives 
\be
\sigma = 
a^2  \epsilon \nabla^{-2}\dot{\psi} - {\psi \over  H} \, ,
\ee
and substituting into the traceless equation and applying $\nabla^2$ we finally obtain
\begin{align}
{1 \over a^3} {\ud \over \ud t}( a^3 \epsilon \dot{\psi})  - \epsilon  {\nabla^2 \over a^2} \psi  =0 \, ,
\end{align}
which is the well-known EOM of curvature perturbation $\psi= -\mathcal{R}$.

Now we move to the open case. Adding the extrinsic curvature combination
\be
E_{\mu \nu} = \mathcal{E}_{\mu \nu} + \Gamma  ( K_{\mu \nu} - K P_{\mu \nu} ) \sqrt{-g^{00}}\, ,
\ee
and fixing the background to FLRW modifies the system of equations to
\begin{align}
\label{eq:00-excurv}
E_{00} &=  {2\over a^2} \nabla^2 \psi + {2 H\over a^2} \nabla^2 \sigma -  6 H \left(\dot{\psi} + H A \right) - 2 \dot{H}A + 2(M^2 -  \Gamma H )A  \, , 
\\ 
\label{eq:0i-excurv}
E_{0i} &= 2\partial_i \left(\dot{\psi} + H A \right) \, , 
\\
\label{eq:ij-excurv}
E_{ij} &=  2 a^2 \left[(3 H + \Gamma) \left(\dot{\psi} + H A \right) +  \partial_t\left(\dot{\psi} +  H A \right)\right]\delta_{ij} 
+ \mathcal{D}_{ij}\left(\psi - A  + H \sigma + \dot{\sigma} + \Gamma   \sigma \right) \, .
\end{align}
We observe that these equations obey the linearized scalar \textit{deformed identity}
\be \label{eq:scalar_deformed_identity}
\partial^j E_{ij} = a^2 \left[ (3 H +\Gamma) E_{0i} + \dot{E}_{0i} \right] \, ,
\ee
which is the scalar part of the deformed identity \eqref{eq:deformed_identity}. 
Due to this identity, the number of independent scalar equations is reduced from four to three, corresponding to the three independent variables $A$, $\psi$ and $\sigma$.

Let us find the EOM of $\psi$ by solving this system of equations. \eqref{eq:00-excurv} and \eqref{eq:0i-excurv} imply 
\be \label{eq:nabla_sigma_open}
\nabla^2 \sigma = - {\nabla^2 \psi \over H} + a^2 \frac{\epsilon+\gamma}{c_{\rm s}^2} \dot{\psi} \, ,
\ee
where $\gamma \equiv \Gamma/H$, 
and substituted into the traceless part of \eqref{eq:ij-excurv} yields
the desired EOM for $\psi$:
\begin{align}
\label{eq:homogeneous_open}
\mathcal{E}_\psi \equiv 
\ddot\psi + \bigg\{ (3+\gamma)H + \partial_t \bigg[ \log \bigg( \frac{2\epsilon+2\gamma}{c_{\rm s}^2} \bigg) \bigg] \bigg\} \dot\psi
- c_{\rm s}^2 \frac{\nabla^2\psi}{a^2} = 0
\, .
\end{align}
Including noise contributions, \eqref{eq:nabla_sigma_open} becomes
\be
\nabla^2 \sigma = - {\nabla^2 \psi \over  H} 
+  a^2 \frac{\epsilon+\gamma}{c_{\rm s}^2} \dot{\psi} 
+ {a^2 \over 2H} \Xi_{00} 
+ {a^2 \over 2} \bigg( 3 -  \frac{\epsilon+\gamma}{c_{\rm s}^2} \bigg) \nabla^{-2}\partial^i\Xi_{0i} \, ,
\ee
and we obtain the Langevin equation
\be
\begin{aligned}
\mathcal{E}_\psi &= 
a^{-2} \bigg( 3 - \frac{\epsilon+\gamma}{c_{\rm s}^2} \bigg)^{-1}
\Bigg\{
{3\over 2} \mathcal{T}^{ij}\Xi_{ij} + {\partial^i\Xi_{0i} \over 2H}  
\\
& \hspace{10em}
- \bigg[ {\ud \over \ud t} + (H+\Gamma) \bigg]
\left[ {a^2 \over 2H} \Xi_{00} + {a^2 \over 2} \bigg( 3 - \frac{\epsilon+\gamma}{c_{\rm s}^2} \bigg) \nabla^{-2}\partial^i\Xi_{0i} \right] \Bigg\}
\, .
\end{aligned}
\ee

\subsection{Vectors} \label{app:vectors}

The linear EOM for vector perturbations in the standard EFT of inflation can be expressed in terms of the vector combination
\be
V_i \equiv  \dot{w}_i - {1\over a} v_i \, ,
\ee
as
\begin{align}
& \mathcal{E}_{0i} =  {1 \over 2 }  \nabla^2 V_i  \, ,\\
& \mathcal{E}_{ij} =  a^2 \left(\dot{V}_{(i,j)} + 3 H V_{(i,j)} \right) \, ,
\end{align}
and satisfy the \textit{vector identity}
\be
\partial^j \mathcal{E}_{ij}= a^2 \left(\dot{\mathcal{E}}_{0i} + 3H \mathcal{E}_{0i} \right)\, ,
\ee
i.e.~the vector part of the divergence equation. 

Adding the extrinsic curvature combination yields the following system:
\begin{align}
& E_{0i} =  {1 \over 2 }  \nabla^2 V_i  \, ,\\
& E_{ij} =  a^2 \left[\dot{V}_{(i,j)} + (3 H +\Gamma) V_{(i,j)} \right] \, ,
\end{align}
which satisfy the linearized vector \textit{deformed identity}
\be \label{eq:vector_deformed_identity}
\partial^j E_{ij} = a^2 \left[ (3 H +\Gamma) E_{0i} + \dot{E}_{0i} \right] \, ,
\ee
which, again, is the vector part of \eqref{eq:deformed_identity}. This identity reduces the number of independent equations from four to two to match the two independent vector degrees in the combination $V_i$.

In the absence of noise, vector perturbations can be set to zero everywhere in space. With noise, the $ij$ component of the noise field is constrained to satisfy the identity \eqref{eq:vector_deformed_identity}, while the $0i$ component sources the (physical) vector field
\be
\nabla^2 V_i = 2 \Xi^V_{0i} \, .
\ee
For vanishing vector perturbations it is necessary to assume vanishing vector noise fluctuations.

\section{The homogeneous solution} 
\label{app:homogeneous}

In this section, we derive the homogeneous solution of \eqref{eq:curv_pertur}. First, we move to the conformal time $\tau$:
\be
\psi_{,\tau \tau} + \left\{ \left( 2 + {\gamma } \right) {a_{,\tau} \over a} 
+ 2\, \partial_\tau \left[\log \left({\sqrt{2\epsilon + 2\gamma} \over c_{\rm s}}   \right)\right] \right\} \psi_{,\tau}
- c_{\rm s}^2 \nabla^2 \psi  =0 \, ,
\ee
and observe that it takes the form of the Mukhanov-Sasaki equation:
\be
\psi_{,\tau \tau} + 2 (\log z)_{,\tau}\psi_{,\tau} -c_{\rm s}^2 \nabla^2 \psi = 0 \, ,
\ee
with 
\be
z \equiv a^{1+\gamma/2} { \sqrt{2\epsilon + 2\gamma} \over c_{\rm s} } \, .
\ee
A further redefinition $u \equiv z \psi$  transforms it in a form suitable for quantization:
\be
\label{eq:mode-eq}
u_{,\tau \tau} +  \bigg( c_{\rm s}^2k^2 - 
{z_{,\tau \tau} \over z} \bigg) u = 0 \, .
\ee
Up to $\mathcal{O}\left[ \epsilon, \eta, (\log c_{\rm s})', (\log \gamma)' \right]$ corrections, where a prime denotes differentiation with respect to the $e$-folding number, \eqref{eq:mode-eq} has as solutions the usual Hankel functions:
%
%
\begin{equation}
u(k,\tau) = \sqrt{-\tau} \Big[ c_1\mathrm{H}_\nu^{(1)}(-c_{\rm s}k\tau) + c_2 \mathrm{H}_\nu^{(2)}(-c_{\rm s}k\tau) \Big] \, ,
\end{equation}
with $\nu = (3+\gamma)/2$.
%
Consider a non-zero $\gamma $ for $\tau \rightarrow - \infty$ which is equivalent to extending the validity of the EFT at arbitrarily high energies. Then, imposing a Bunch-Davies initial state, we can fix $c_1 = \sqrt{\pi} e^{i(\nu+1/2) \pi/2}/2$ and $c_2 = 0$, and find the solution 
as
%
%
\begin{align}
|\psi|
& =
\frac{\sqrt{\pi}}{2} \frac{c_{\rm s}}{\sqrt{2\epsilon+2\gamma}} H^{1+\gamma/2} (-\tau)^{(3+\gamma)/2}
\mathrm{H}_{(3+\gamma)/2}^{(1)}(-c_{\rm s}k\tau)
\nonumber\\
& \underset{\tau\rightarrow 0^-}{\longrightarrow}
\frac{2^{(1+\gamma)/2}}{\sqrt{\pi}} \frac{H^{1+\gamma/2}}{c_{\rm s}^{(1+\gamma)/2} \sqrt{2\epsilon+2\gamma}}
\Gamma \bigg( \frac{3+\gamma}{2} \bigg) k^{-(3+\gamma)/2}
\, .
\end{align}
Using Stirling's approximation on the $\Gamma$ function, for large dissipation the dominant behaviour of the solution scales as $|\psi|_{\tau\rightarrow 0^-} \sim \sqrt{\gamma}^{\gamma}$, a result which seems inconsistent because we extend the EFT beyond its validity. A more realistic alternative is to consider that $\gamma$ becomes significant after some finite $\tau_0$ and then to match the general solution with the asymptotic Bunch-Davies solution without dissipation \cite{LopezNacir:2011kk,Creminelli:2023aly}. In doing so, the solution becomes
\begin{align} 
\label{eq:homogeneous_general}
%
\psi
& =
-\sqrt{\frac{\pi}{8\epsilon}} \tilde{c}_{\rm s}H \bigg( \frac{\gamma\omega_0}{c_{\rm s}k} \bigg)^{-\gamma/2}
e^{-i\left(1-\frac{\tilde{c}_{\rm s}}{c_{\rm s}}\right)\gamma\omega_0} e^{- i\frac{\pi}{4}\gamma}
\sqrt{\frac{c_{\rm s}}{\tilde{c}_{\rm s}}} 
\nonumber\\
& \quad
\times
\bigg[ 1 + i \frac{8(c_{\rm s}-\tilde{c}_{\rm s}) - \gamma\tilde{c}_{\rm s}(6+\gamma)}{8\tilde{c}_{\rm s}\gamma\omega_0} \bigg]
(-\tau)^{\nu} \mathrm{H}_\nu^{(1)}(-c_{\rm s}k\tau)
\, ,
\end{align}
where $\tilde{c}_{\rm s}$ and $c_{\rm s}$ denote respectively the sound speeds before and after $\tau_0$, and $\omega_0 \equiv - {\tau_0  c_{\rm s} k / \gamma }$. We have also assumed  $\omega_0 \gg 1$, or that we can assign the Bunch-Davies solution sufficiently far in the past. For small $\gamma$ we recover the standard behaviour of the mode function in the closed theory, while for large $\gamma$ the solution is exponentially suppressed by the dissipation rate $\gamma$:
\be
|\psi|_{\tau \rightarrow 0^-} = 
e^{-\gamma/2} \omega_0^{-\gamma/2} \frac{\gamma^2H}{8\omega_0}
\sqrt{\frac{\tilde{c}_{\rm s}}{c_{\rm s}^2\epsilon k^3}}
\, .
\ee

\end{appendix}

\addcontentsline{toc}{section}{References}
\bibliographystyle{JHEP}
\bibliography{eftoi}

@article{Cheung:2007st,
    author = "Cheung, Clifford and Creminelli, Paolo and Fitzpatrick, A. Liam and Kaplan, Jared and Senatore, Leonardo",
    title = "{The Effective Field Theory of Inflation}",
    eprint = "0709.0293",
    archivePrefix = "arXiv",
    primaryClass = "hep-th",
    reportNumber = "IC-2007-032",
    doi = "10.1088/1126-6708/2008/03/014",
    journal = "JHEP",
    volume = "03",
    pages = "014",
    year = "2008"
}

@article{Hongo:2018ant,
    author = "Hongo, Masaru and Kim, Suro and Noumi, Toshifumi and Ota, Atsuhisa",
    title = "{Effective field theory of time-translational symmetry breaking in nonequilibrium open system}",
    eprint = "1805.06240",
    archivePrefix = "arXiv",
    primaryClass = "hep-th",
    reportNumber = "RIKEN-iTHEMS-Report-18, KOBE-COSMO-18-05, RIKEN-ITHEMS-REPORT-18",
    doi = "10.1007/JHEP02(2019)131",
    journal = "JHEP",
    volume = "02",
    pages = "131",
    year = "2019"
}

@article{Salcedo:2024smn,
    author = "Salcedo, Santiago Agui and Colas, Thomas and Pajer, Enrico",
    title = "{The open effective field theory of inflation}",
    eprint = "2404.15416",
    archivePrefix = "arXiv",
    primaryClass = "hep-th",
    doi = "10.1007/JHEP10(2024)248",
    journal = "JHEP",
    volume = "10",
    pages = "248",
    year = "2024"
}

@article{Liu:2018kfw,
    author = "Liu, Hong and Glorioso, Paolo",
    title = "{Lectures on non-equilibrium effective field theories and fluctuating hydrodynamics}",
    eprint = "1805.09331",
    archivePrefix = "arXiv",
    primaryClass = "hep-th",
    reportNumber = "MIT-CTP/5018; EFI-18-8, MIT-CTP-5018, EFI-18-8",
    doi = "10.22323/1.305.0008",
    journal = "PoS",
    volume = "TASI2017",
    pages = "008",
    year = "2018"
}

@article{Crossley:2015evo,
    author = "Crossley, Michael and Glorioso, Paolo and Liu, Hong",
    title = "{Effective field theory of dissipative fluids}",
    eprint = "1511.03646",
    archivePrefix = "arXiv",
    primaryClass = "hep-th",
    reportNumber = "MIT-CTP-4734",
    doi = "10.1007/JHEP09(2017)095",
    journal = "JHEP",
    volume = "09",
    pages = "095",
    year = "2017"
}

@article{Malik:2008im,
    author = "Malik, Karim A. and Wands, David",
    title = "{Cosmological perturbations}",
    eprint = "0809.4944",
    archivePrefix = "arXiv",
    primaryClass = "astro-ph",
    doi = "10.1016/j.physrep.2009.03.001",
    journal = "Phys. Rept.",
    volume = "475",
    pages = "1--51",
    year = "2009"
}

@article{LopezNacir:2011kk,
    author = "Lopez Nacir, Diana and Porto, Rafael A. and Senatore, Leonardo and Zaldarriaga, Matias",
    title = "{Dissipative effects in the Effective Field Theory of Inflation}",
    eprint = "1109.4192",
    archivePrefix = "arXiv",
    primaryClass = "hep-th",
    reportNumber = "SLAC-PUB-14995",
    doi = "10.1007/JHEP01(2012)075",
    journal = "JHEP",
    volume = "01",
    pages = "075",
    year = "2012"
}

@article{Delacretaz:2021qqu,
    author = "Delacr\'etaz, Luca V. and Gout\'eraux, Blaise and Ziogas, Vaios",
    title = "{Damping of Pseudo-Goldstone Fields}",
    eprint = "2111.13459",
    archivePrefix = "arXiv",
    primaryClass = "hep-th",
    reportNumber = "CPHT-RR097.112021",
    doi = "10.1103/PhysRevLett.128.141601",
    journal = "Phys. Rev. Lett.",
    volume = "128",
    number = "14",
    pages = "141601",
    year = "2022"
}

@article{Armas:2021vku,
    author = "Armas, Jay and Jain, Akash and Lier, Ruben",
    title = "{Approximate symmetries, pseudo-Goldstones, and the second law of thermodynamics}",
    eprint = "2112.14373",
    archivePrefix = "arXiv",
    primaryClass = "hep-th",
    doi = "10.1103/PhysRevD.108.086011",
    journal = "Phys. Rev. D",
    volume = "108",
    number = "8",
    pages = "086011",
    year = "2023"
}

@article{Baggioli:2023tlc,
    author = "Baggioli, Matteo and Bu, Yanyan and Ziogas, Vaios",
    title = "{U(1) quasi-hydrodynamics: Schwinger-Keldysh effective field theory and holography}",
    eprint = "2304.14173",
    archivePrefix = "arXiv",
    primaryClass = "hep-th",
    reportNumber = "CPHT-RR017.042023",
    doi = "10.1007/JHEP09(2023)019",
    journal = "JHEP",
    volume = "09",
    pages = "019",
    year = "2023"
}

@article{Salcedo:2024nex,
    author = "Salcedo, Santiago Agui and Colas, Thomas and Pajer, Enrico",
    title = "{An Open Effective Field Theory for light in a medium}",
    eprint = "2412.12299",
    archivePrefix = "arXiv",
    primaryClass = "hep-th",
    doi = "10.1007/JHEP03(2025)138",
    journal = "JHEP",
    volume = "03",
    pages = "138",
    year = "2025"
}

@article{Creminelli:2023aly,
    author = "Creminelli, Paolo and Kumar, Soubhik and Salehian, Borna and Santoni, Luca",
    title = "{Dissipative inflation via scalar production}",
    eprint = "2305.07695",
    archivePrefix = "arXiv",
    primaryClass = "hep-th",
    doi = "10.1088/1475-7516/2023/08/076",
    journal = "JCAP",
    volume = "08",
    pages = "076",
    year = "2023"
}

@article{Salcedo:2025ezu,
    author = {Salcedo, Santiago Ag{\"u}{\'\i} and Colas, Thomas and Dufner, Lennard and Pajer, Enrico},
    title = "{An open system approach to gravity}",
    eprint = "2507.03103",
    archivePrefix = "arXiv",
    primaryClass = "hep-th",
    doi = "10.1007/JHEP02(2026)241",
    journal = "JHEP",
    volume = "02",
    pages = "241",
    year = "2026"
}

@article{Caldeira:1983uj,
  author = {Caldeira, A. O. and Leggett, A. J.},
  title = {Path Integral Approach to Quantum Brownian Motion},
  journal = {Physica A},
  volume = {121},
  pages = {587--616},
  year = {1983},
  doi = {10.1016/0378-4371(83)90013-4}
}

@article{Keldysh:1964ud,
    author = "Keldysh, L. V.",
    title = "{Diagram Technique for Nonequilibrium Processes}",
    doi = "10.1142/9789811279461_0007",
    journal = "Sov. Phys. JETP",
    volume = "20",
    pages = "1018--1026",
    year = "1965"
}

@article{Schwinger:1960qe,
    author = "Schwinger, Julian S.",
    title = "{Brownian motion of a quantum oscillator}",
    doi = "10.1063/1.1703727",
    journal = "J. Math. Phys.",
    volume = "2",
    pages = "407--432",
    year = "1961"
}

@article{Sieberer:2015hba,
    author = {Sieberer, L. M. and Chiocchetta, A. and Gambassi, A. and T{\"a}uber, U. C. and Diehl, S.},
    title = "{Thermodynamic Equilibrium as a Symmetry of the Schwinger-Keldysh Action}",
    eprint = "1505.00912",
    archivePrefix = "arXiv",
    primaryClass = "cond-mat.stat-mech",
    doi = "10.1103/PhysRevB.92.134307",
    journal = "Phys. Rev. B",
    volume = "92",
    number = "13",
    pages = "134307",
    year = "2015"
}

@article{Haehl:2016pec,
    author = "Haehl, Felix M. and Loganayagam, R. and Rangamani, Mukund",
    title = "{Schwinger-Keldysh formalism. Part I: BRST symmetries and superspace}",
    eprint = "1610.01940",
    archivePrefix = "arXiv",
    primaryClass = "hep-th",
    doi = "10.1007/JHEP06(2017)069",
    journal = "JHEP",
    volume = "06",
    pages = "069",
    year = "2017"
}

@article{Feynman:1963fq,
    author = "Feynman, R. P. and Vernon, Jr., F. L.",
    editor = "Brown, L. M.",
    title = "{The Theory of a general quantum system interacting with a linear dissipative system}",
    doi = "10.1016/0003-4916(63)90068-X",
    journal = "Annals Phys.",
    volume = "24",
    pages = "118--173",
    year = "1963"
}

@article{Glorioso:2017fpd,
    author = "Glorioso, Paolo and Crossley, Michael and Liu, Hong",
    title = "{Effective field theory of dissipative fluids (II): classical limit, dynamical KMS symmetry and entropy current}",
    eprint = "1701.07817",
    archivePrefix = "arXiv",
    primaryClass = "hep-th",
    reportNumber = "MIT-CTP-4860, EFI-17-2",
    doi = "10.1007/JHEP09(2017)096",
    journal = "JHEP",
    volume = "09",
    pages = "096",
    year = "2017"
}

@article{Lau:2024mqm,
    author = "Lau, Pak Hang Chris and Nishii, Kanji and Noumi, Toshifumi",
    title = "{Gravitational EFT for dissipative open systems}",
    eprint = "2412.21136",
    archivePrefix = "arXiv",
    primaryClass = "hep-th",
    reportNumber = "KOBE-COSMO-24-06, UT-Komaba/24-11",
    doi = "10.1007/JHEP02(2025)155",
    journal = "JHEP",
    volume = "02",
    pages = "155",
    year = "2025"
}

@article{Christodoulidis:2025ymc,
    author = "Christodoulidis, Perseas",
    title = "{Emergent structures in open EFTs}",
    eprint = "2509.13284",
    archivePrefix = "arXiv",
    primaryClass = "hep-th",
    month = "9",
    year = "2025"
}

@article{Gumrukcuoglu:2013nza,
    author = {G{\"u}mr{\"u}k{\c{c}}{\"u}o{\u{g}}lu, A. Emir and Hinterbichler, Kurt and Lin, Chunshan and Mukohyama, Shinji and Trodden, Mark},
    title = "{Cosmological Perturbations in Extended Massive Gravity}",
    eprint = "1304.0449",
    archivePrefix = "arXiv",
    primaryClass = "hep-th",
    doi = "10.1103/PhysRevD.88.024023",
    journal = "Phys. Rev. D",
    volume = "88",
    number = "2",
    pages = "024023",
    year = "2013"
}

@article{deRham:2014naa,
    author = "de Rham, Claudia and Heisenberg, Lavinia and Ribeiro, Raquel H.",
    title = "{On couplings to matter in massive (bi-)gravity}",
    eprint = "1408.1678",
    archivePrefix = "arXiv",
    primaryClass = "hep-th",
    doi = "10.1088/0264-9381/32/3/035022",
    journal = "Class. Quant. Grav.",
    volume = "32",
    pages = "035022",
    year = "2015"
}

@article{Mukhanov:1990me,
    author = "Mukhanov, Viatcheslav F. and Feldman, H. A. and Brandenberger, Robert H.",
    title = "{Theory of cosmological perturbations. Part 1. Classical perturbations. Part 2. Quantum theory of perturbations. Part 3. Extensions}",
    reportNumber = "BROWN-HET-796, BROWN-HET-800, BROWN-HET-780",
    doi = "10.1016/0370-1573(92)90044-Z",
    journal = "Phys. Rept.",
    volume = "215",
    pages = "203--333",
    year = "1992"
}

@article{Berera:1995ie,
    author = "Berera, Arjun",
    title = "{Warm inflation}",
    eprint = "astro-ph/9509049",
    archivePrefix = "arXiv",
    reportNumber = "PSU-TH-159",
    doi = "10.1103/PhysRevLett.75.3218",
    journal = "Phys. Rev. Lett.",
    volume = "75",
    pages = "3218--3221",
    year = "1995"
}

@article{Arnowitt:1962hi,
    author = "Arnowitt, Richard L. and Deser, Stanley and Misner, Charles W.",
    title = "{The Dynamics of general relativity}",
    eprint = "gr-qc/0405109",
    archivePrefix = "arXiv",
    doi = "10.1007/s10714-008-0661-1",
    journal = "Gen. Rel. Grav.",
    volume = "40",
    pages = "1997--2027",
    year = "2008"
}

@article{Kaplanek:2025moq,
    author = "Kaplanek, Greg and Mylova, Maria and Tolley, Andrew J.",
    title = "{Gauging Open EFTs from the top down}",
    eprint = "2512.17089",
    archivePrefix = "arXiv",
    primaryClass = "hep-th",
    reportNumber = "Imperial/TP/2025/AJT/1",
    month = "12",
    year = "2025"
}

@article{Gong:2023kpe,
    author = "Gong, Jinn-Ouk and Mylova, Maria and Sasaki, Misao",
    title = "{New shape of parity-violating graviton non-Gaussianity}",
    eprint = "2303.05178",
    archivePrefix = "arXiv",
    primaryClass = "hep-th",
    reportNumber = "APCTP-Pre2023-003, YITP-23-23",
    doi = "10.1007/JHEP10(2023)140",
    journal = "JHEP",
    volume = "10",
    pages = "140",
    year = "2023"
}

@article{Christodoulidis:2024ric,
    author = "Christodoulidis, Perseas and Gong, Jinn-Ouk and Lin, Wei-Chen and Mylova, Maria and Sasaki, Misao",
    title = "{New shape for cross-bispectra in Chern-Simons gravity}",
    eprint = "2409.09935",
    archivePrefix = "arXiv",
    primaryClass = "hep-th",
    doi = "10.1088/1475-7516/2025/01/037",
    journal = "JCAP",
    volume = "01",
    pages = "037",
    year = "2025"
}

@article{Calzetta:1995ys,
    author = "Calzetta, E. and Hu, B. L.",
    title = "{Quantum fluctuations, decoherence of the mean field, and structure formation in the early universe}",
    eprint = "gr-qc/9505046",
    archivePrefix = "arXiv",
    reportNumber = "IASSNS-HEP-95-38, UMD-PP-95-083",
    doi = "10.1103/PhysRevD.52.6770",
    journal = "Phys. Rev. D",
    volume = "52",
    pages = "6770--6788",
    year = "1995"
}

@article{Boyanovsky:2015tba,
    author = "Boyanovsky, D.",
    title = "{Effective field theory during inflation: Reduced density matrix and its quantum master equation}",
    eprint = "1506.07395",
    archivePrefix = "arXiv",
    primaryClass = "astro-ph.CO",
    doi = "10.1103/PhysRevD.92.023527",
    journal = "Phys. Rev. D",
    volume = "92",
    number = "2",
    pages = "023527",
    year = "2015"
}

@article{Green:2024hbw,
    author = "Green, Daniel and Gupta, Kshitij and Huang, Yiwen",
    title = "{A Goldstone boson equivalence for inflation}",
    eprint = "2403.05274",
    archivePrefix = "arXiv",
    primaryClass = "hep-th",
    doi = "10.1007/JHEP09(2024)117",
    journal = "JHEP",
    volume = "09",
    pages = "117",
    year = "2024"
}

\end{document}